# Magneto-Tunable Thermal Diode Based on Bulk Superconductor

Poonam Rani[1,†], Masayuki Mashiko[1,†], Keisuke Hirata[2], Ken-ichi Uchida[2,3], Yoshikazu Mizuguchi[1,*]

[1]*Department of Physics, Tokyo Metropolitan University, Hachioji, Japan*
[2]*Department of Advanced Materials Science, Graduate School of Frontier Sciences, The University of Tokyo, Kashiwa, Japan*
[3]*National Institute for Materials Science (NIMS), Tsukuba, Japan*
*†These authors contributed equally to this work.*
*E-mail: mizugu@tmu.ac.jp*

**Abstract**

Thermal diode is a growing technology and important for active thermal flow control. Since the theoretical designing of thermal diode in 2004, various kinds of solid-state thermal diodes have been theoretically and experimentally investigated. Here, we report on the observation of thermal rectification in bulk-size superconductor-normal metal junctions. High-purity (5N) wires of Pb and Al are soldered, and thermal conductivity ($\kappa$) of the junctions is measured in two different directions of the heat flow, forward ($\kappa_\text{F}$) and reverse ($\kappa_\text{R}$) directions. Thermal rectification ratio ($\kappa_\text{F} / \kappa_\text{R}$) of 1.75 is obtained at $T \sim 5.2$ K with $H = 400$ Oe. The merit of the Pb-Al junction is a large difference of $\kappa$ in an order of several hundred W m$^{-1}$ K$^{-1}$ and magneto-tunability of the working temperature.

**1. Introduction**

Thermal management is one of the technologies crucial for creating new application and improving the performance of electronic devices.[1–4] Among the thermal management technologies, thermal switches, which achieve a large change in thermal conductivity ($\kappa$), and thermal diodes, which enable thermal rectification under temperature difference, are essential for active heat control. Materials and devices for thermal switches and diodes have been actively studied,[5–8] and superconducting materials are good candidate for thermal switching because of a large change in $\kappa$ through the superconducting transition, by changing temperature ($T$) and/or magnetic field ($H$), while the working temperature is limited to below the transition temperature ($T_\text{SC}$).[9] Using pure-element superconductors, a large magneto-thermal switching (MTS) with a MTS ratio (MTSR), which is calculated as MTSR $(T, H) = [\kappa(T, H) - \kappa(T, H = 0 \text{ Oe})]/\kappa(T, H = 0 \text{ Oe})$, can be obtained due to the large change in electronic thermal conductivity ($\kappa_{el}$). In the superconducting states, electrons form Cooper pairs, and the carrier thermal transport is suppressed. For high-purity Nb and Pb samples, MTSR of 650% and 2000% were observed, respectively.[10,11] Furthermore, nonvolatility of MTS can be achieved using phase-separated superconducting alloys because of flux trapping, which results in the suppression of bulkiness of superconducting states of the lower-$T_{SC}$ regions and higher $\kappa_{el}$; nonvolatile MTS has been observed in Sn-Pb and In-Sn solders.[12,13] Nonvolatile MTS appears also in a type-II superconductor Nb in the mixed states due to the changes in the lattice thermal conductivity ($\kappa_{lat}$) in it mixed states; at higher temperature the fluxes also affect $\kappa_{el}$.[14] Nonvolatile MTS at a higher temperature can be achieved using high-$T_{SC}$ superconductors like $MgB_2$,[15] although the switching ratio is still small (<20% in comparison between initial and demagnetized $\kappa$).

After the theoretical design of thermal diodes,[16-19] the thermal diodes using superconductors have been proposed in 2013.[20,21] The concept of superconductor-based thermal diodes (SC thermal diode) is based on the junction made of superconductor and normal metal or the use of Josephson junction, which is composed of superconductor-insulator-superconductor junction. In general, the thermal rectification effect can be attributed to asymmetric interfacial thermal resistance between dissimilar materials. This mechanism, along with its extensions to various material systems, has been comprehensively discussed in a recent review article.[22] Zhang et al. recently demonstrated a sustainable all-solid elastocaloric cooling device that exploits non-reciprocal heat transfer to enhance efficiency and reduce energy consumption.[23] If efficient SC thermal diodes can be obtained and easily used in various cryogenic applications, efficiency of cryogenic devices will be highly improved owing to active thermal control. However, most studies on SC thermal diodes have been based on simulation and investigation of nano-scale devices at low temperatures (lower than 1 K),[24–26] and the experimental observation of SC thermal diode effect in a bulk-size materials has not been reported so far. Inspired by recent works on thermal rectification and related applications, we fabricated a thermal diode based on superconductors. Here, we show that the bulk superconductor-normal conductor junction made of high-purity wires of Pb (5N purity) and Al (5N purity) exhibits clear thermal rectification. In this paper, effective thermal conductivity for the junction ($\kappa^*$) is measured where the $\kappa^*$ was calculated by assuming that the sample shape is a uniform wire with a diameter of 0.5 mm. The highest thermal rectification ratio (TRR), defined as TRR = $\kappa^*_F/\kappa^*_R$ where $\kappa^*_F$ and $\kappa^*_R$ are $\kappa^*$ measured in the forward and reverse direction, respectively, reaching 1.75 was observed at $T = 5.2$ K under $H = 400$ Oe. Furthermore, the Pb-Al diode exhibits a large difference in effective thermal conductivity (Δ$κ$*) between forward and reverse heat directions; the largest Δ$κ$* observed in this study is 270 W m$^{-1}$ K$^{-1}$. The large Δ$κ$* would be the advantage of this material in designing practical applications. In addition, the Pb-Al thermal diode can work with a small temperature difference, and the working (TRR-maximum) temperature is systematically tunable by an applied magnetic field. The Pb–Al thermal diode could open new pathway to the application on cryogenic thermal regulation for superconducting quantum circuits and sensors, space instrumentation where passive, low-loss thermal control is essential, and energy-efficient refrigeration or waste-heat harvesting systems. These applications would highlight how superconductivity-driven rectification can contribute to sustainable thermal management and next-generation phononic or caloritronic devices.

## 2. Fabrication of SC Thermal Diode

The design strategy for SC thermal diode based on the Pb and Al wires is summarized in **Figure 1**. To obtain larger TRR and Δ$κ$*, a larger change in $κ$* is preferred. The total thermal resistance ($W_{tot}$) is given by the summation of thermal resistance at the Pb wire ($W_{Pb}$), the Al wire ($W_{Al}$), and the joint ($W_{joint}$). For example, $W_{Pb}$ is calculated using $W = L/κS$ where $L$ and $S$ are the distance from one thermometer to the joint and cross-sectional area, respectively. Therefore, we need a metal with high $κ$, and Al (5N) was used for this study. The joint between Pb and Al wires were made using Sn60-Pb40 solder. As reported in Ref. 12, Sn-Pb solder is a phase-separated composite, and $T_{sc}$ of the solder is 7.2 K, same as pure Pb. As discussed in discussion, the thermal resistance of the solder part is high, and its change in thermal conductivity at $T_c$ is limited. Furthermore, the temperature gradient is basically made in the wire parts. Therefore, the high thermal resistance at the joint has less affection on the thermal rectification properties.

Since thermal diodes work under the presence of temperature deference,[18–20] $T_{SC}$ of Pb should be in between $T$ of two thermometers, $T$ (high) and $T$ (low), to get a higher TRR. The setup for measurements using thermal transport option (TTO) on Physical Property Measurement System (PPMS-Dynacool, Quantum Design) is displayed in **Figure 2**. By reversing the orientation of the sample, the forward and reverse measurements have been performed. Here, we define the forward direction as shown in **Figure 1**; Pb is on the hot side for forward measurements. In the ideal case of the forward measurements where the center temperature is $T_{sc}$ of Pb, both Pb and Al are in normal conducting states with higher $κ$; $T_{SC}$ of Al is lower than the examined $T$ range. In contrast, in the reverse setup, the Al part is normal conducting, but the Pb part becomes superconductive with low $\kappa$. The strategy for the designing SC thermal diode is quite simple as explained here, but the experimental observation has been challenging because the reduction of $\kappa$ below $T_{SC}$ is generally broad even in pure-element superconductors. As reported in Ref. 11, even in the Pb-5N wire, the drop of $\kappa$ in $\kappa$-$T$ is broad at around $T_{SC}$. when the applied $H$ is perpendicular to the heat flow direction. We recently, however, found that the $\kappa$-$T$ for the 5N-Pb wire exhibits a quite sharp reduction when the applied $H$ is parallel to the heat flow direction (**Figure 3 (a)**).[27] This anomalous temperature dependence enables the observation of SC thermal diode effect with the presence of small temperature difference ($\Delta T$) between the two thermometers. **Figure 3(b)** shows the $\kappa$-$T$ for Al-5N wire measured under various $H$. There are positive magneto-thermal resistance effects, and the temperature dependence is also not negligible. Therefore, thermal rectification is also observed at $T$ far from the $T_{SC}$, which is consistent with the conventional thermal diode design.[18,19] because of high $\kappa$ for the fabricated Pb-Al diode, we measured long samples as shown in **Figure 2(b)**.

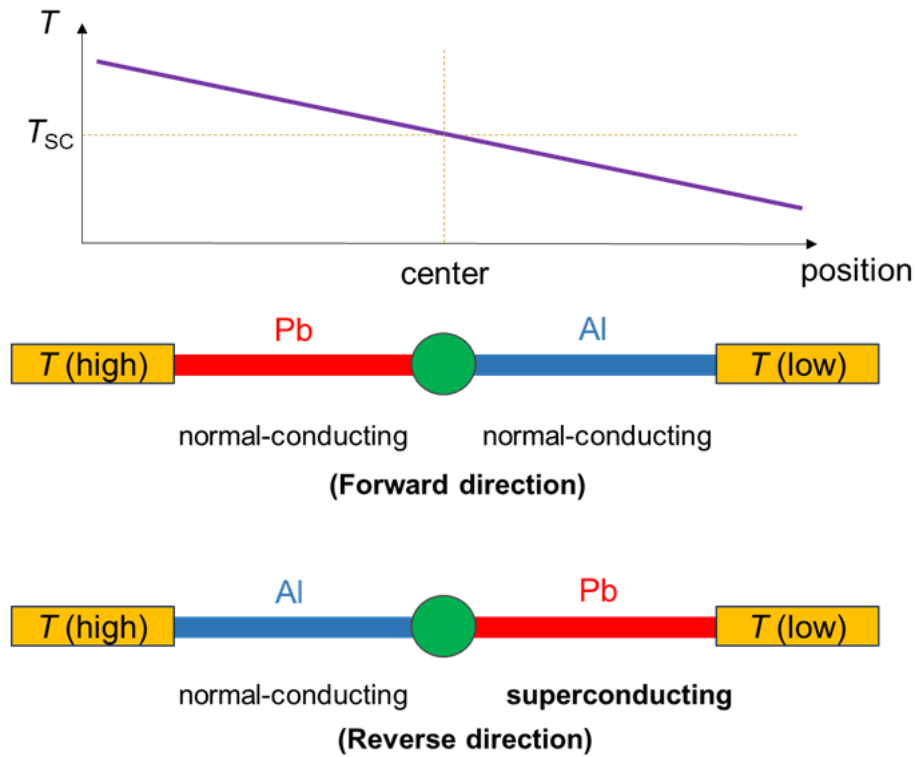

**Figure 1.** Design strategy for Pb-Al thermal diode.

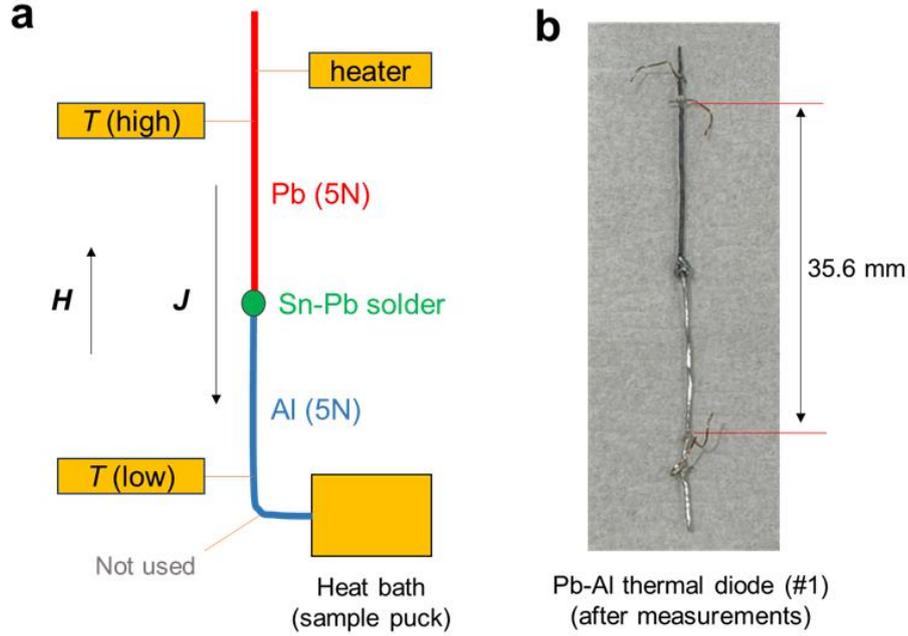

**Figure 2.** (a) Sample setup for TTO measurements. (b) Photo of the fabricated Pb-Al thermal diode (sample #1). The length ratio of Pb:Al is close to 1:1 (Pb50%-Al50%).

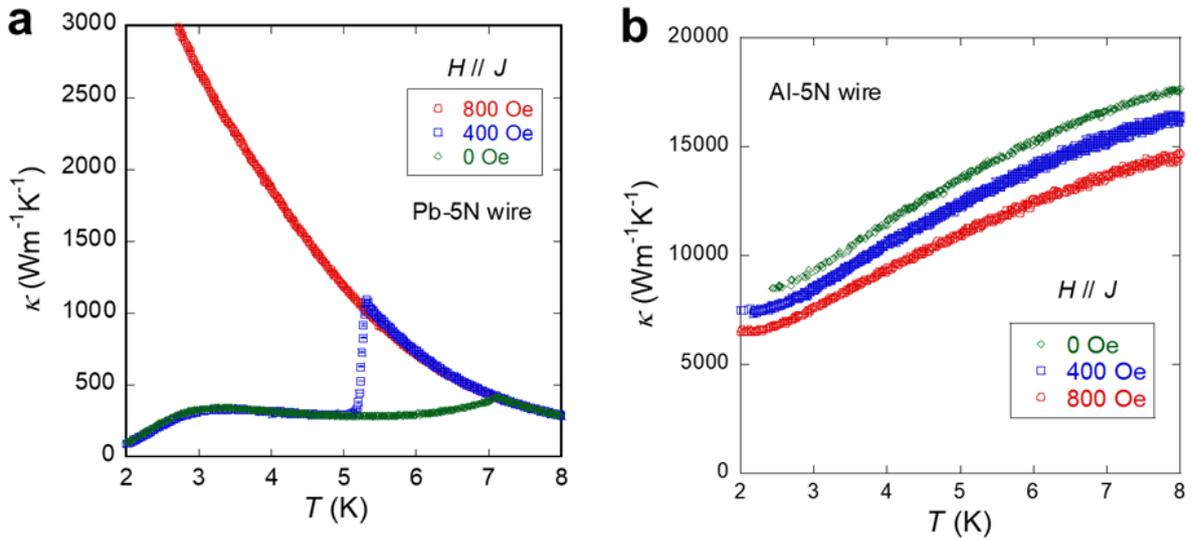

**Figure 3.** $T$ dependence of $\kappa$ under magnetic field ($H$ = 0, 400, 800 Oe) of the used Pb-5N wire (a) and Al-5N wire (b). $H$ was applied parallel to the heat flow. The original data of (a) has been published in Ref. 27.

## 3. Results and Discussion

In **Figure 4**, we show both the forward and reverse results ($\kappa^*$-$T$); the $T$ of the horizontal axis is average $T$ between $T$ (high) and $T$ (low). A clear difference in the $\kappa^*$-drop temperature is observed at $H$ = 400 Oe as shown in **Figure 4(a)**. The observation of the large difference between $\kappa^*_F$ and $\kappa^*_R$ suggests efficient thermal rectification at the temperature. At $H = 0$ Oe, the $\Delta\kappa^*$ is small in the whole $T$ range while the sift of $T_{SC}$ is observed. Clear difference appears at $H = 200$ Oe, which suggests the thermal rectification. At $H = 600$ Oe, the temperature dependence changes, and the $\kappa^*$ far from $T_{SC}$ exhibits $\Delta\kappa^*$. This is caused by the thermal rectification originated from the different gradient of $\kappa$-$T$ particularly in the normal states of Pb and Al as shown in **Figure 3**. At $H = 800$ Oe, no superconducting transition is observed, and small rectification is seen due to the above-mentioned mechanism. In the $\kappa^*$-$T$ at $H = 600$ and 800 Oe, the normal-state $\kappa^*$ ($T > T_{sc}$) decreases with decreasing temperature. Because of the large increase in $\kappa$ in Pb at lower temperature, higher $\kappa^*$ is simply expected due to high $\kappa$ of both the Pb and Al wires. In Table 1, $\kappa$, $W$, and $\kappa^*$ (average of $\kappa^*_F$ and $\kappa^*_R$) for the Pb and Al wires at $H = 800$ Oe and the calculated $W$ at the Sn-Pb joint are summarized. Here, the length of the Pb and Al wires between is assumed as 17.8 mm. The estimated joint $W$ at $T = 3.0$ K is 1.55 times larger than that $W$ at $T = 4.0$ K, which is consistent with the temperature evolution of $\kappa$ of Sn60-Pb40 solder; $\kappa = 29$ and 44 W m$^{-1}$ K$^{-1}$ at $T = 3.0$ and 4.0 K, respectively (1.52 times larger at $T = 3.0$ K).[12] The discussion above clearly suggests that the resulting $\kappa^*$ is affected by the joint $W$, and the improvement of the joint condition of the fabricated Pb-Al thermal diode can result in a higher $\kappa^*$. Although the absolute value of $\kappa^*$ is affected by the joint $W$, the working temperature of the thermal diode and TRR are basically determined by the properties of the wire parts.

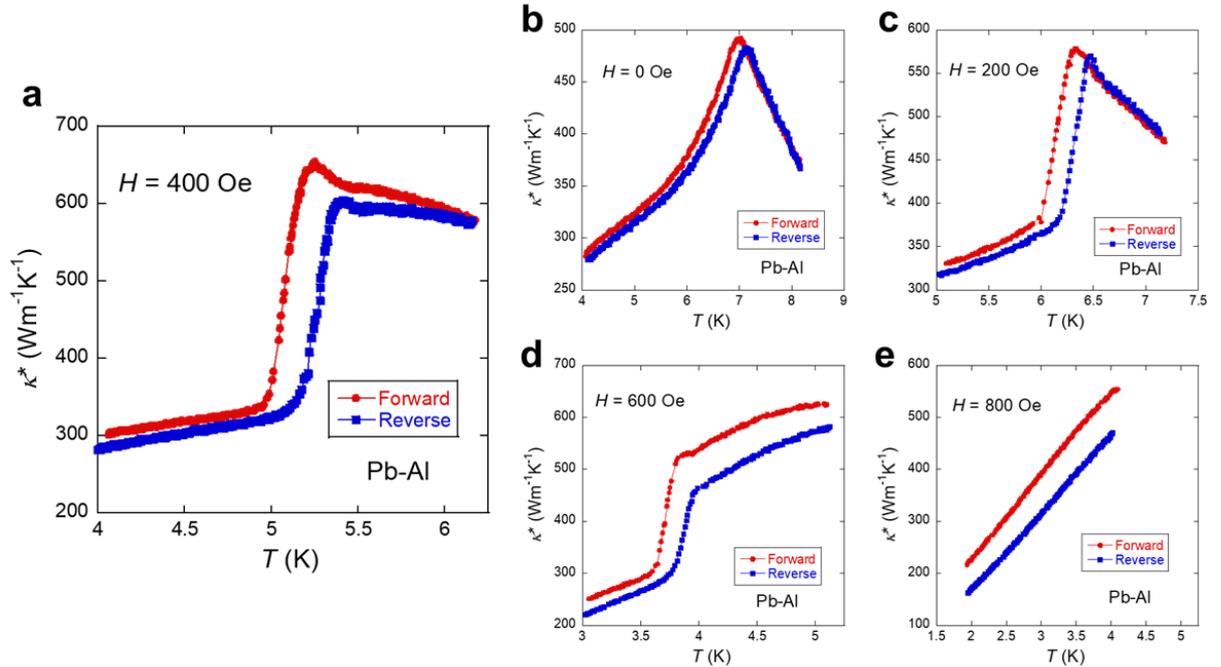

**Figure 4.** $T$ dependence of $\kappa^*$ under magnetic field of $H =$ (b) 0, (c) 200, (a) 400, (d) 600, (e) 800 Oe for the Pb-Al diode (sample #1) on the forward and reverse setups.

**Table 1.** Data at $H = 800$ Oe and calculation of joint thermal resistance. The $\kappa$ data of Sn60-Pb40 solder was taken from Ref. 12.

| $T$ (K) | $\kappa$ (W m$^{-1}$ K$^{-1}$) | | $W$ (K W$^{-1}$) | | $\kappa^*$ (W m$^{-1}$ K$^{-1}$) Average of $\kappa^*_F$ and $\kappa^*_R$ | Calculated $W_{joint}$ (K W$^{-1}$) (Sn60-Pb40 solder) | $\kappa$ (W m$^{-1}$ K$^{-1}$) (Sn60-Pb40 solder) |
|---|---|---|---|---|---|---|---|
| | Pb | Al | Pb | Al | | | |
| 3.0 | 2700 | 7550 | 33.6 | 12.0 | 350 | 473 | 29 [Ref. 12] |
| 4.0 | 1880 | 9350 | 48.2 | 9.7 | 500 | 305 | 44 [Ref. 12] |

Using the $\kappa^*$-$T$ data shown in **Figure 4**, the $T$ dependence of $\Delta\kappa^*$ and TRR are estimated and plotted in **Figure 5**. With increasing $H$, the peak temperature of $\Delta\kappa^*$ and TRR clearly shifts to a lower temperature, which proofs that the largest $\Delta\kappa^*$ and TRR can be achieved using the sharp drop of $\kappa$ near the superconducting transition of Pb. The observed TRR is not the highest among the experimentally studied thermal diodes,[7] the observed $\Delta\kappa^*$ is the largest among them to the best of our knowledge. The studies aiming a higher TRR have been focusing on the materials with low $\kappa_{lat}$; hence, the total $\kappa$ of the target material was relatively low. The present study uses high-$\kappa$ metals, and the resulting rectification $\Delta\kappa^*$ is quite large. We consider that the thermal rectification with a large $\Delta\kappa^*$ will open new pathway of cryogenic thermal management.

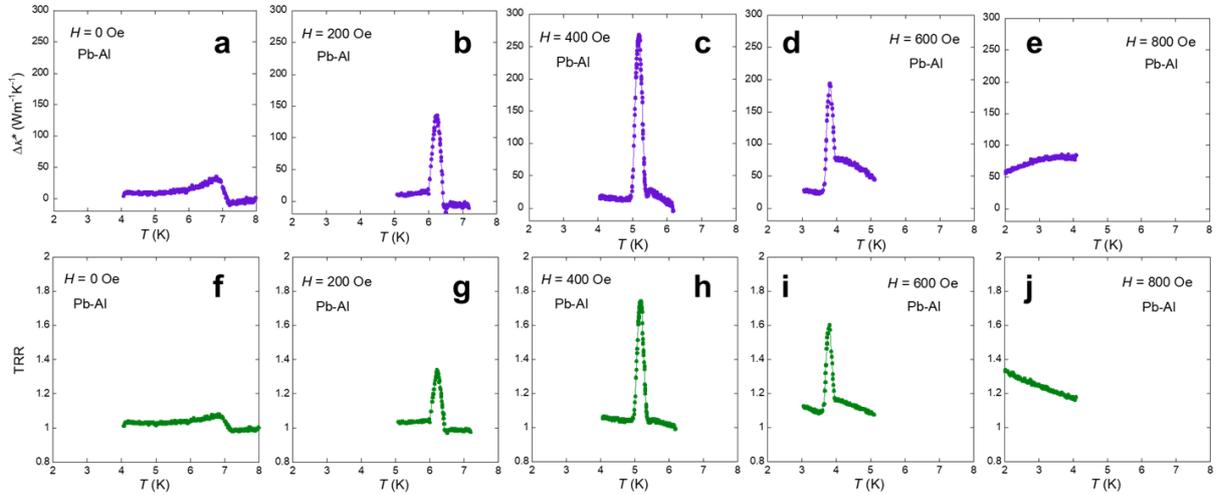

**Figure 5.** $T$ dependence of $\Delta\kappa^*$ under magnetic field of $H =$ (a) 0, (b) 200, (c) 400, (d) 600, (e) 800 Oe for the Pb-Al diode (sample #1). $T$ dependence of TRR under magnetic field of $H =$ (f) 0, (g) 200, (h) 400, (i) 600, (j) 800 Oe for the Pb-Al diode (sample #1).

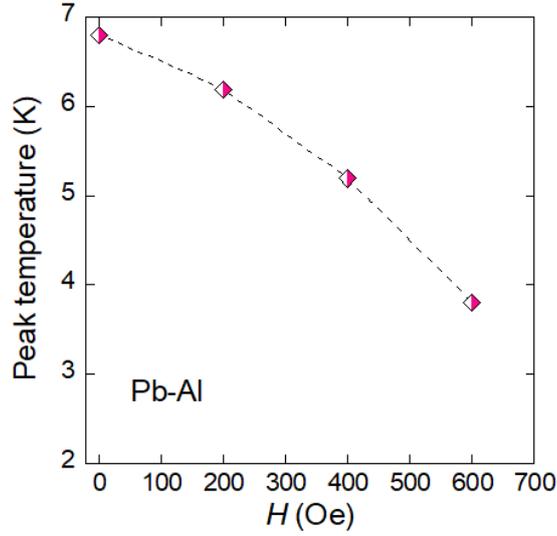

**Figure 6.** *H* dependence of peak temperature where the TRR becomes maximum.

From the peak position of the TRR in TRR-*T*, we estimate the field dependence of the peak temperature of TRR (**Figure 6**). The peak temperature indicates the temperature at which the thermal rectification is maximized. Therefore, the peak temperature demonstrates the best working temperature of the SC thermal diode. As shown in **Figure 6**, the working temperature is systematically tuned by applied magnetic field, which is related to critical field of Pb. Because we expect that the SC thermal diode is used at low temperatures under potential temperature variation (fluctuation), tunability of working temperature by changing the applied magnetic field is useful for maximizing the thermal rectification. Importantly, *H* needed for the control is lower than 1000 Oe because the critical field of Pb is 800 Oe (*T* = 0 K). The tunability with low magnetic field enables the use of SC diode as a simple component (structure) in cryogenic devices. To confirm reproducibility of the results shown above, four different samples were investigated and showed similar behavior. $\kappa^*$-*T*, $\Delta\kappa^*$, and TTR for sample #2 are summarized in **Figure S1** (Supporting Information).

In this study, we measured $\kappa^*$ using the target temperature rise of 10% of the average temperature. At temperature near the superconducting transition, however, the resulting temperature rise and $\Delta T$ slightly changed. In a conventional way of the estimation of the thermal rectification and TRR is the comparison of $\kappa^*$ or heat flow density (*J*) with the same $\Delta T$.[8] Therefore, to confirm the emergence of the thermal rectification in the present Pb-Al thermal diode, many $\kappa^*$ data were taken, and the $\kappa$ having the same *T* and $\Delta T$ were selected and compared between the forward and reverse conditions (**Table 2**). From the data, we confirm that the similar TRR can be reproduced under the same $\Delta T$.

**Table 2.** Effective thermal conductivity and TRR measured with similar $\Delta T$ values between the forward and reverse measurements.

| $T$ (K) | $\Delta T$ (K) | $\kappa^*_F$ (W m$^{-1}$ K$^{-1}$) | $\kappa^*_R$ (W m$^{-1}$ K$^{-1}$) | TRR |
|---|---|---|---|---|
| 5.34 | 0.30 | 543 | 580 | 0.94 |
| 5.12 | 0.34 | 512 | 340 | 1.51 |
| 5.10 | 0.33 | 489 | 332 | 1.47 |

To investigate the effect of the length ratio of Pb and Al, the samples of Pb40%-Al60% and Pb60%-Al40% where the Pb length between the two thermometers is 40% and 60% of the total length, respectively. In **Figure S2** (Supporting Information), the temperature dependence of $\kappa^*$, $\Delta\kappa^*$, and TRR for the Pb40%-Al60% and Pb60%-Al40% is summarized. The observed TRR is slightly lower than that observed for the Pb50%-Al50% sample mainly discussed in this work. As shown in **Figure S2(b)**, the $\Delta\kappa^*$ for Pb40%-Al60% is larger than that for Pb50%-Al50%, which is caused by low thermal resistance of the Al part. Therefore, $\Delta\kappa^*$ can be optimized by changing the Pb-to-Al ratio.

## 4. Conclusion

We fabricated superconductor thermal diodes using high-purity (5N) polycrystalline wires of superconducting Pb and normal-conducting Al. Under **H** // **J**, Pb wire shows a sharp drop of $\kappa$ at $T_{SC}$, and the thermal rectification was clearly observed for the Pb-Al junction, which works as a thermal diode. The largest TRR and $\Delta\kappa^*$ observed at $H$ = 400 Oe and $T \sim 5.2$ K were 1.75 and 270 W m$^{-1}$ K$^{-1}$, respectively. The temperature where the TRR becomes largest can be tuned by $H$, which suggests that the thermal rectification performance is easily maximized according to the environment. The observation of thermal rectification in simple bulk superconductor-based junction will be useful for further development of cryogenic thermal management devices because of the tunability of the diode size and working temperature.

## 5. Experimental Section/Methods

The Pb (5N purity) and Al (5N purity) polycrystalline wires with a diameter of 0.5 mm were purchased from The Nilaco Corporation. To make a joint, we used Sn-Pb (flux-cored) commercial solders (Sn60-Pb40). Measurements of $\kappa$ with four-terminal methods were performed on Physical Property Measurement System (PPMS-Dynacool, Quantum Design) using the thermal transport option (TTO). For the $\kappa$ measurements, the terminals were fabricated using Ag paste and Cu wires with a diameter of 0.2 mm, and the field direction (**H** // **J**) was controlled by changing the sample setup. The typical measurement period was 5-10 s, and the temperature sweep speed for $\kappa^*$-$T$ was 0.05 K min$^{-1}$. To obtain relatively large temperature difference between two thermometers, the target temperature rise was fixed to 10%; we confirmed that this condition results in a good waveform in the $\kappa^*$ measurements. For the $\kappa$-$T$ measurements shown in **Figure 3**, the target temperature rise of 3% was used.

**Acknowledgements**

The authors thank M. Yoshida, H. Arima, T. Ichikawa, A. Yamashita, and Y. Oikawa for discussion on the thermal transport and diode effect. This work was partly supported by JST-ERATO (No.: JPMJER2201) and TMU research funds for young scientists.

**Data Availability Statement**

All the data presented in this article can be provided by reasonable requests to the corresponding author.

**References**

[1] Jia, J., Li, S., Chen, X., & Shigesato, Y., Emerging Solid–State Thermal Switching Materials, Adv. Funct. Mater. 34, 2406667 (2024). https://doi.org/10.1002/adfm.202406667.

[2] Li, N., Ren, J., Wang, L., Zhang, G., Hanggi, P., & Li, B., Phononics: Manipulating heat flow with electronic analogs and beyond, Rev. Mod. Phys. 84, 1045 (2012). https://doi.org/10.1103/RevModPhys.84.1045.

[3] Wehmeyer, G., Yabuki, T., Monachon, C., Wu, J., & Dames, C., Thermal diodes, regulators, and switches: Physical mechanisms and potential applications, Appl. Phys. Rev. 4, 041304 (2017). https://doi.org/10.1063/1.5001072.

[4] Zheng, Q., Hao, M., Miao, R., Schaadt, J., & Dames, C., Advances in thermal conductivity for energy applications: a review, Prog. Energy 3, 012002 (2021). https://doi.org/10.1088/2516-1083/abd082.

[5] Kimling, J., Gooth, J., & Nielsch, K., Spin-dependent thermal transport perpendicular to the planes of Co/Cu multilayers, Phys. Rev. B 91, 144405 (2015). https://doi.org/10.1103/PhysRevB.91.144405.

[6] Nakayama, H., Xu, B., Iwamoto, S., Yamamoto, K., Iguchi, R., Miura, A., Hirai, T., Miura, Y., Sakuraba, Y., Shiomi, J., & Uchida, K., Above-room-temperature giant thermal conductivity switching in spintronic multilayers, Appl. Phys. Lett. 118, 042409 (2021). https://doi.org/10.1063/5.0032531.

[7] Wong, M. Y., Tso, C. Y., Ho, T. C., & Lee, H. H., A review of state of the art thermal diodes and their potential applications, Int. J. Heat Mass Transfer 164, 120607 (2021). https://doi.org/10.1016/j.ijheatmasstransfer.2020.120607.

[8] Hirata, K., Matsunaga, T., Singh, S., Matsunami, M., & Takeuchi, T., High-Performance Solid-State Thermal Diode Consisting of $Ag_2$(S,Se,Te), J. Electron. Mater. 49, 2895 (2020), https://doi.org/10.1007/s11664-020-07964-8.

[9] Arima, H., Murakami, T., Rani, P., & Mizuguchi, Y., Magneto-thermal switching using superconducting metals and alloys, Sci. Technol. Adv. Mater., https://doi.org/10.1080/14686996.2025.2506978.

[10] Yoshida, M., Kasem, M. R., Yamashita, A., Uchida, K., & Mizuguchi, Y., Magneto-thermal-switching properties of superconducting Nb, Appl. Phys. Express 16, 033002 (2023), https://doi.org/10.35848/1882-0786/acc3dd.

[11] Yoshida, M., Arima, H., Yamashita, A., Uchida, K., & Mizuguchi, Y., Large magneto-thermal-switching ratio in superconducting Pb wires, J. Appl. Phys. 134, 065102 (2023), https://doi.org/10.1063/5.0159336

[12] Arima, H., Kasem, M. R., Sepehri-Amin, H., Ando, F., Uchida, K., Kinoshita, Y., Tokunaga, M., & Mizuguchi, Y., Observation of nonvolatile magneto-thermal switching in superconductors, Commun. Mater. 5, 34(1-8) (2024), https://doi.org/10.1038/s43246-024-00465-9.

[13] Rani, P., Murakami, T., Watanabe, Y., Sepehri-Amin, H., Arima, H., Yamashita, A., Mizuguchi, Y., Nonvolatile magneto-thermal switching driven by vortex trapping in commercial In-Sn solder, Appl. Phys. Express 18, 033001(1-6) (2025), https://doi.org/10.35848/1882-0786/adb6ee.

[14] Kes, P. H., van der Veeken, J. P. M. & de Kierk, D., Thermal Conductivity of Niobium in the Mixed State, J. Low Temp. Phys. 18, 355 (1975), https://doi.org/10.1007/BF00118165.

[15] Arima, H. & Mizuguchi, Y., Nonvolatile Magneto-Thermal Switching in $MgB_2$, J. Phys. Soc. Jpn. 92, 103702 (2023), https://doi.org/10.7566/JPSJ.92.103702.

[16] Li, B., Wang, L., & Casati, G., Thermal Diode: Rectification of Heat Flux, PRL 93 (18), 184301 (2004), https://doi.org/10.1103/PhysRevLett.93.184301.

[17] Li, B., Lan, J. & Wang, L. Interface Thermal Resistance between Dissimilar Anharmonic Lattices, PRL 95, 104302 (2005), https://doi.org/ 10.1103/PhysRevLett.95.104302.

[18] Peyrard, M., The design of a thermal rectifier, EPL 76, 49 (2006), https://doi.org/10.1209/epl/i2006-10223-5.

[19] Hu, B., He, D., Yang, L., & Zhang, Y., Thermal rectifying effect in macroscopic size, Phys. Rev. E 74, 060201 (2006), https://doi.org/10.1103/PhysRevE.74.060201.

[20] Giazotto, F., & Bergeret, F. S., Thermal rectification of electrons in hybrid normal metal-superconductor nanojunctions, Applied Physics Letters 103, 242602 (2013), https://doi.org/10.1063/1.4846375.

[21] Martínez-Pérez, M. J. & Giazotto, F, Efficient phase-tunable Josephson thermal rectifier, Appl. Phys. Lett. 102, 182602 (2013), https://doi.org/10.1063/1.4804550.

[22] Chen, J., Xu, X., Zhou, J. & Li, B., Interfacial thermal resistance: Past, present, and future, Rev. Mod Phys 94, 025002 (2022), https://doi.org/10.1103/RevModPhys.94.025002.

[23] Zhang, J., Shen, X., Chen, M., Luo, W., Wei, B., Luo, T., Li, B. & Zhu, G., Sustainable all-solid elastocaloric cooler enabled by non-reciprocal heat transfer, nature sustainability 8, 650 (2025), https://doi.org/10.1038/s41893-025-01552-6

[24] Martínez-Pérez, M. J., Fornieri, A., & Giazotto, F., Rectification of electronic heat current by a hybrid thermal diode, Nat. Nanotech. 10, 303 (2015), https://doi.org/10.1038/NNANO.2015.11.

[25] Fornieri, A. & Giazotto, F., Towards phase-coherent caloritronics in superconducting circuits, Nat. Nanotech. 12, 944 (2017), https://doi.org/10.1038/NNANO.2017.204.

[26] Antola, F., Braggio, F., De Simoni, G., & Giazotto, F., Tunable thermoelectric superconducting heat pipe and diode, Supercond. Sci. Technol. 37, 115023 (2024), https://doi.org/10.1088/1361-6668/ad7d40.

[27] Rani, P., Watanabe, Y., Shiga, T., Sakuraba, Y., Takeda, H., Yamashita, M., Uchida, K., Yamashita, A., & Mizuguchi, Y., Huge anisotropic magneto-thermal switching in high-purity polycrystalline compensated metals, arXiv:2506.00427, https://doi.org/10.48550/arXiv.2506.00427.

**Supporting Information**

To figure

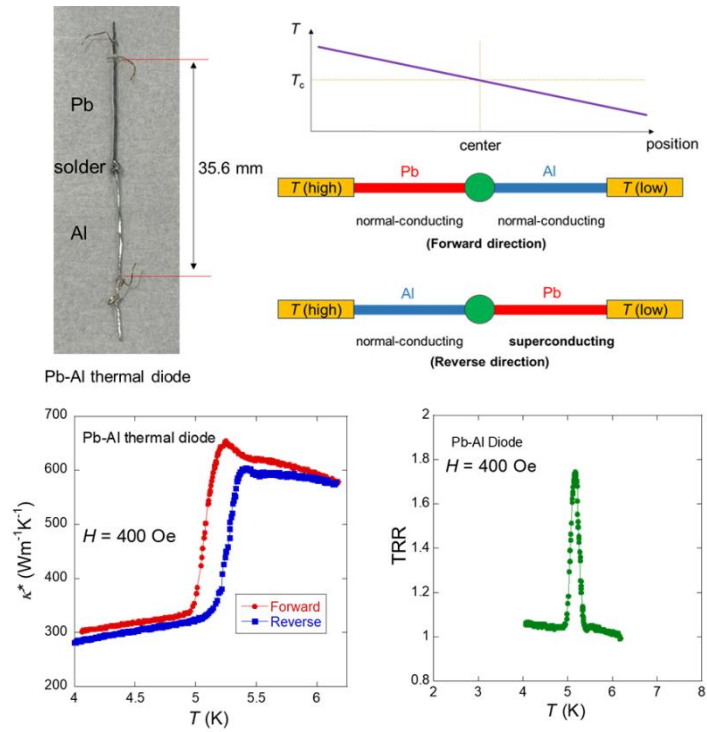

Thermal diode based on bulk superconducting materials is developed. Using the sharp change in thermal conductivity and its temperature dependence of Pb near the superconducting transition, Pb-Al junction works as a thermal diode with thermal rectification ratio (TRR) of 1.75. The working temperature can be tuned by applied magnetic field.

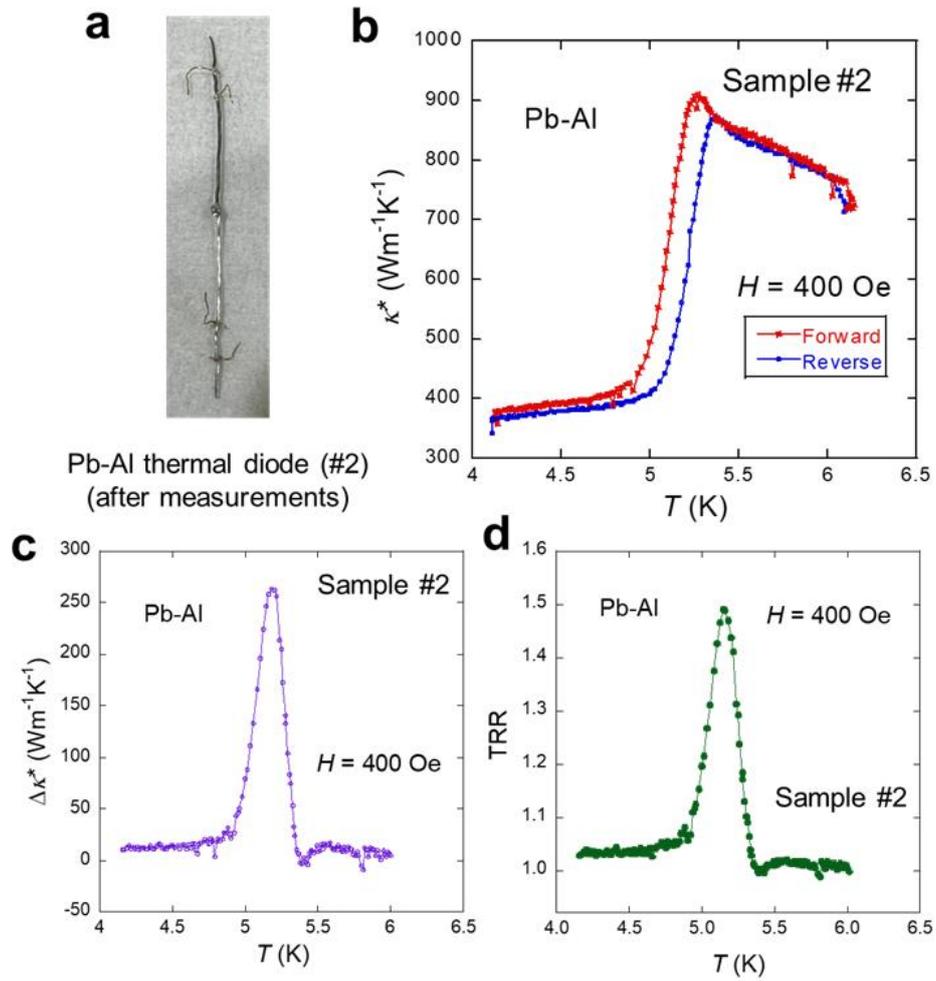

**Figure S1**. (a) Photo of the Pb-Al diode (sample #2). (b–d) Temperature dependence of (b) $\kappa^*$, (c) $\Delta\kappa^*$, and (d) TRR for sample #2 measured at $H$ = 400 Oe.

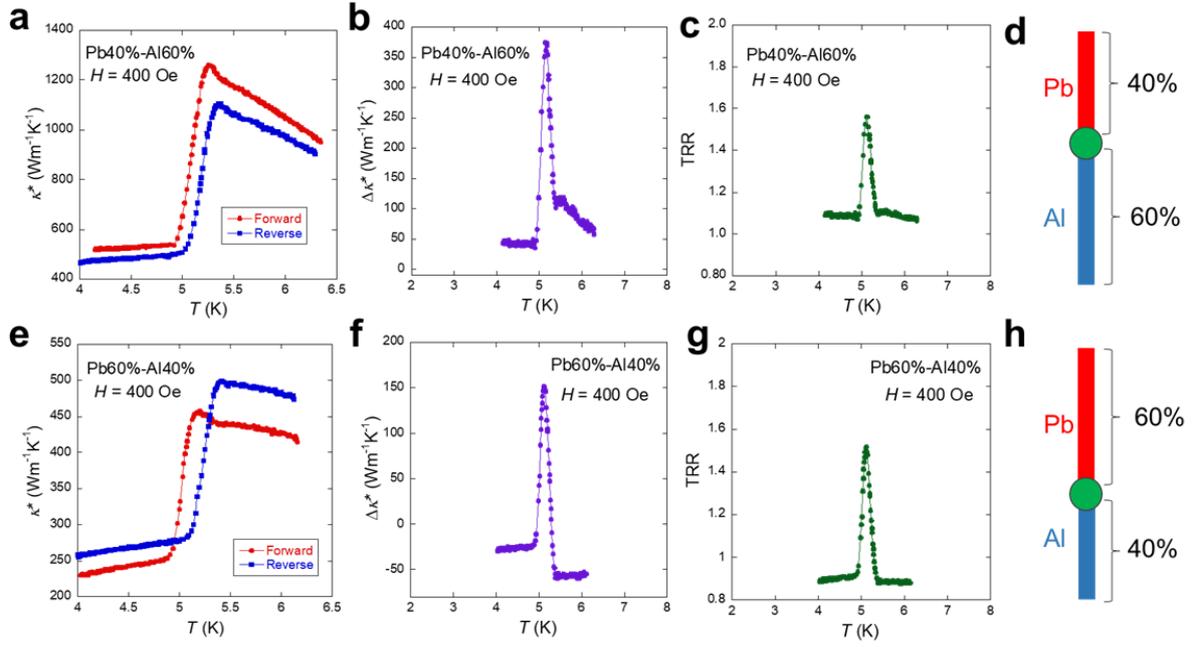

**Figure S2**. (a–c) Temperature dependence of (a) $\kappa^*$, (b) $\Delta\kappa^*$, and (c) TRR for the sample with Pb40%-Al60% ratio. (d) Schematic image of Pb-Al ratio in the Pb40%-Al60% junction. (e–g) Temperature dependence of (d) $\kappa^*$, (e) $\Delta\kappa^*$, and (f) TRR for the sample with Pb60%-Al40% ratio. The data were measured at $H$ = 400 Oe. (g) Schematic image of Pb-Al ratio in the Pb60%-Al40% junction.